\documentclass[12pt]{article}
\usepackage[english]{babel}
\usepackage[pdftex]{color}
\usepackage{amsmath}
\usepackage{graphicx}
\usepackage{hyperref}
\usepackage{amsmath}
\usepackage{amsfonts}
\usepackage{amssymb}

\begin{document}

\title{ \textbf{{}On Composite fields approach to Gribov copies elimination in Yang-Mills theories\thanks{Talk presented at SQS'13,  29 July – 03 August, 2013, at JINR, Dubna, Russia}}}
\author{\textsc{Alexander A. Reshetnyak\thanks{%
reshet@ispms.tsc.ru}} \\
Laboratory of Computer-Aided Design of Materials, Institute of \\
Strength Physics and Materials Science, 634021 Tomsk, Russia, \\
Tomsk State Pedagogical University, 634061 Tomsk, Russia}
\date{}
\maketitle

\begin{abstract}
We suggest a method of introducing the Gribov--Zwanziger horizon functional,
$H$, for Yang--Mills theories by using the composite fields technique: $\sigma
(\phi )=H$. A\ different form of the same horizon functional in gauges $\chi
$ and $\chi ^{\prime }$ is taken into account via (gauged) field-dependent
BRST transformations connecting quantum Yang--Mills actions in these gauges.
We introduce generating functionals of Green's functions with composite
fields and derive the corresponding Ward identities. A study of gauge
dependence shows that the effective action in Yang--Mills theories with the
composite field $H$ does not depend on the gauge on the extremals determined
by the Yang--Mills fields $\phi $ alone.
\end{abstract}

\section{Introduction}

The concept of BRST symmetry, expressing gauge invariance via a special
one-parameter global supersymmetry \cite{brst}, is a crucial instrument in
the quantum description of a gauge theory in view of the fact that the known
fundamental interactions are described in terms of Yang--Mills type gauge
theories \cite{books}. This concept underlies the success of perturbative
calculations at high energies and the numerical study in lattice gauge
theory \cite{lattice}, \cite{lattice2}. It also provides a strong evidence
for the interaction of quarks and gluons, correctly described by a
non-Abelian gauge theory known as QCD.

The problem of gauge-fixing with the help of a differential condition (like
the Coulomb or Landau gauges), as it was first shown by V.~Gribov \cite%
{Gribov} and then studied by I.~Singer \cite{Singer} in Yang--Mills (YM)
theories, cannot be correctly realized within the Faddeev--Popov (FP)
procedure \cite{FP}, even perturbatively, for the entire spectrum of the
momenta distribution for gauge fields, $A_{\mu }^{a}$, in the deep IR
region, due to an infinitely large number of discrete gauge copies, emerging
outside the so-called first Gribov region, $\Omega (A)$. A solution of this
problem can be found by an introduction to the quantum action, constructed
by the Faddeev--Popov procedure, of a special horizon functional, $H(A)$
\cite{Zwanziger1}, \cite{Zwanziger2}, known as the Gribov--Zwanziger (GZ)
theory, which was constructed only for the Landau gauge and is not
BRST-invariant.

Until recently, the study of the complete GZ theory has been carried out
almost entirely within the Landau gauge (see, e.g., \cite{sorella1} and the
references therein), whereas a restriction to $\Omega (A)$ in the path
integral has been made for YM theories in the approximation being quadratic
in the fields and using the covariant \cite{Sorella} and maximal Abelian
gauges \cite{maxabg}. At the same time, there exists a significant
arbitrariness in the choice of admissible gauges, which is partially related
to the choice of a reference frame, see, e.g., \cite{DeWitt}. Thus, it is
well known that Green's functions depend on the choice of a gauge; however,
this dependence has a special structure, such that it should be absent in
physical quantities like the $S$-matrix. The contemporary study of
gauge-dependence and unitarity in the Lorentz-covariant quantum description
of gauge models is based on BRST symmetry. Therefore, any violation of BRST
invariance may result in a gauge-dependent and non-unitary $S$-matrix. A
consideration of the first problem within the concept of \emph{soft BRST
symmetry breaking} in YM and general gauge theories \cite{llr}, \cite{lrr},
within the BV quantization scheme \cite{BV}, has revealed a requirement for
a special transformation of the gauge variation, $\delta M$, for the BRST
symmetry breaking term, $M$, in order to provide the gauge-independence of
the effective action (EA) on the mass shell. Recently, it was shown \cite%
{Reshetnyak} that this requirement for $\delta M$ is always fulfilled within
a class of general gauge theory with \emph{soft breaking of BRST symmetry}
based on the concept of field-dependent BRST symmetry transformations,
introduced for YM theories in \cite{ll1} to relate the quantum FP action
given in a fixed gauge to the one in different gauge, and used to determine
the GZ horizon functional for the GZ theory in the $R_{\xi }$-family of the
gauges. This result solves the problem of gauge-independence for gauge
models with soft BRST symmetry breaking, provided that a model in any fixed
gauge reference frame, in addition to the quantum BV action, is supplied by
a functional $M$ whose form also changes in another gauge by the rule
suggested in \cite{llr}, \cite{lrr}. However, unitarity for YM theories,
studied in \cite{KugoOjima}, has met with an obstacle for the resulting
quantum theory with such an introduction of the soft BRST symmetry breaking
term. In this work, we intend to propose a different method of introducing
the GZ horizon functional, $H(A)$, into the path integral of YM theories: on
the basis of composite fields \cite{compf}. \cite{LO}, \cite{LOR}. Thus, our
purpose is to introduce soft BRST symmetry breaking terms into the FP
quantum action by using composite fields, and then, to obtain the Ward
identities for YM theories with composite fields and investigate the problem
of gauge dependence.

The paper is organized as follows. In Section~\ref{YMtheory}, we remind the
key issues of the soft BRST symmetry breaking in YM theories and derive the
Ward identity for the EA, which provides a basic result involving the
variation of the EA under a variation of the gauge fermion. In Section~\ref%
{GZtheory} we apply this result to the GZ theory. The use of the composite
field technique for an incorporation of the Gribov horizon functional, $H(A)$%
, into the path integral and derivation of the Ward identities, together
with a description of the gauge-dependence problem, is considered in Section~%
\ref{CFmethod}.


\section{Soft BRST Symmetry Breaking in Yang-Mills Theories}

\label{YMtheory} 

The (extended by antifields $\phi _{A}^{\ast }$) generating functional of
Green's functions, $Z_{M}=Z_{M}(J ,\phi ^{\ast })$, for a YM theory with
a soft BRST symmetry breaking term, $M=M(\phi ,\phi ^{\ast })$, [$M(\phi
,0)=m(\phi )$] is determined, along the lines of \cite{llr}, by a path
integral, depending on sources $J_{A}$ for a total set of fields, $\phi ^{A}$%
, complete with the classical (gauge) $A_{\mu }^{a}$, ghost $C^{a}$,
antighost $\overline{C}{}^{a}$, Nakanishi--Lautrup $B^{a}$ fields, for $%
a=1,\ldots ,\dim SU(N)$, $\mu =0,1,\ldots ,D-1$ in DeWitt's condensed
notations and those of Ref. \cite{llr}, as follows:%
\begin{eqnarray}\label{ZM}
  Z_M &=& \int d\phi \exp\left\{\frac{\imath}{\hbar}\big(S_0(A)+s\psi(\phi)+\phi^*_A s\phi^A+ M + J_A\phi^A\big) \right\}.
\end{eqnarray}
Here, a classical action, $S_{0}(A)$, invariant under gauge transformations,
$\delta A_{\mu }^{a}={D}_{\mu }^{ab}\xi ^{b}$, with a YM covariant
derivative, ${D}_{\mu }^{ab}$, completely antisymmetric $SU(N)$ structure
constants $f^{abc}$ and arbitrary functions $\xi ^{b}$ in Minkowski
space-time, $R^{1,D-1}$, a nilpotent Slavnov variation, $s$, and a gauge
fermion, $\psi $, are given by the relations,
\begin{eqnarray}  \label{gtrYM}
&& S_{0}(A)  =  {-}\frac{1}{4}\int d^{D}x\ F_{\mu \nu }^{a}F^{\mu \nu {}a}%
\texttt{ for } [D_{\mu }^{ab}, F_{\mu \nu }^{a}] = [
 {\delta}^{ab}\partial _{\mu }+f^{acb}A_{\mu }^{c}, {\partial}_{[\mu }A_{\nu]
}^{a}+f^{abc}A_{\mu }^{b}A_{\nu }^{c}]\ ,
\label{clYM} \\
&&  sF(\phi,\phi^*)   =  \frac{\delta F}{\delta\phi^A} s\phi^A, \qquad  s\phi^A = ({D}_{\mu }^{ab}C^b, \textstyle\frac{1}{2}f^{abc}C^bC^c, B^a, 0),
\label{Slav}\\
&&  \psi(\phi) = \overline{C}^a \chi^a(A,B)\texttt{ for the gauge } \chi^a(A,B) = \partial^\mu A^a_\mu + \textstyle\frac{\xi}{2}B^a = 0.\label{rxig}
\end{eqnarray}%
In terms of the operator $s$, the BRST-non-invariance of the bosonic
functionals $M$, $m$ implies $(sM,sm)\neq (0,0)$. For a vanishing $M$, we
deal with a usual path integral and $Z\equiv Z_{0}(J,\phi ^{\ast })$.

The Ward identities for $Z_{M}$ and EA (generating functional of vertex
Green's functions), $\Gamma _{M}=\Gamma _{M}(\phi ,\phi ^{\ast })$,
introduced via a Legendre transformation of $\frac{\hbar }{\imath }\ln Z_{M}$
with respect to $J_{A}$: $\Gamma _{M}$ = $\frac{\hbar }{\imath }\ln
Z_{M}-J\phi $ for $\phi =\frac{\hbar }{\imath }(\delta \ln Z_{M})/(\delta J)$
have the form
\begin{eqnarray}
&&\Big(J_{A}+M_{A}\big({\textstyle\frac{\hbar }{\text{i}}}{\textstyle\frac{%
\delta }{\delta J}},\phi ^{\ast }\big)\Big)\left( \frac{\hbar }{\text{i}}%
\frac{\delta }{\delta \phi _{A}^{\ast }}\ -\ M^{A\ast }\big({\textstyle\frac{%
\hbar }{\text{i}}}{\textstyle\frac{\delta }{\delta J}},\phi ^{\ast }\big)%
\right) Z_{M}=0,  \label{wardZM} \\
&&\frac{\delta \Gamma _{M}}{\delta \phi ^{A}}\frac{\delta \Gamma _{M}}{%
\delta \phi _{A}^{\ast }}\ =\frac{\delta \Gamma _{M}}{\delta \phi ^{A}}{%
\widehat{M}}^{A\ast }+{\widehat{M}}_{A}\frac{\delta \Gamma _{M}}{\delta \phi
_{A}^{\ast }}-{\widehat{M}}_{A}{\widehat{M}}^{A\ast }\ .  \label{WIGammaBV}
\end{eqnarray}%
where the notation $\big(M_{A},M^{A\ast }\big)\big({\textstyle\frac{\hbar }{%
\text{i}}}{\textstyle\frac{\delta }{\delta J}},\phi ^{\ast }\big)\equiv \Big(%
\frac{\delta M}{\delta \phi ^{A}},\frac{\delta M}{\delta \phi _{A}^{\ast }}%
\Big)\big|_{\phi \rightarrow \frac{\hbar }{\text{i}}\frac{\delta }{\delta J}%
} $, and $\big(\widehat{M}_{A},\widehat{M}{}^{A\ast }\big)$ $\equiv $ $\big(%
M_{A},$ $M^{A\ast }\big)\big|_{\phi \rightarrow \widehat{\phi }}$ has been
used in accordance with \cite{llr}, \cite{lrr}, \cite{Reshetnyak}, with
operator-valued  fields
\begin{eqnarray}
 \label{MAG}
{\widehat\phi}{}^A\ = \phi^A + \imath\hbar\,(\Gamma_M^{''-1})^{AB}
\frac{\delta_l}{\delta\phi^B}, \texttt{
with } (\Gamma_M^{''-1})^{AC}(\Gamma_M^{''})_{CB}=\delta^A_{\ B},\ (\Gamma^{''}_M)_{AB}\ =\ \frac{\delta_l}{\delta\phi^A}
\Big(\frac{\delta\Gamma_M}{\delta\phi^B}\Big)
.
\end{eqnarray}%
In obtaining (\ref{wardZM}), (\ref{WIGammaBV}), we have not utilized the
BRST symmetry breaking equation \cite{llr}, $M_{A},M^{A\ast }=0$, In turn,
the result for the gauge-dependence of $Z_{M},\Gamma _{M}$ can be presented
as follows, for $\delta \psi =\frac{\xi }{2}\overline{C}^{a}B^{a}$,
\begin{eqnarray}
&&\delta Z_{M}=\frac{\imath }{\hbar }\Big[\big(J_{A}+M_{A}(\textstyle\frac{%
\hbar }{\imath }\frac{\delta }{\delta J},\phi ^{\ast })\big)\Big(\frac{%
\delta }{\delta \phi _{A}^{\ast }}-\frac{\imath }{\hbar }M^{A\ast }(\frac{%
\hbar }{\imath }\frac{\delta }{\delta J},\phi ^{\ast })\Big)\delta \psi ({%
\textstyle\frac{\hbar }{\imath }}{\textstyle\frac{\delta }{\delta J}}%
)+\delta M(\frac{\hbar }{\imath }\frac{\delta }{\delta J},\phi ^{\ast })\Big]%
Z_{M},  \label{varZMlin} \\
&&\delta \Gamma _{M}\ =\frac{\delta \Gamma _{M}}{\delta \phi ^{A}}{\widehat{F%
}}_{M}^{A}\,\widehat{\delta \Psi }\ -\ {\widehat{M}}_{A}{\widehat{F}}_{M}^{A}%
\widehat{\delta \psi }\ +\ \widehat{\delta M}\ ,  \label{varGMlin} \\
&&\mathtt{for\ }{\widehat{F}}_{M}^{A}\equiv -\frac{\delta }{\delta \phi
_{A}^{\ast }}\ +\ (-1)^{\varepsilon _{B}(\varepsilon _{A}+1)}(\Gamma
_{M}^{^{\prime \prime }-1})^{BC}\Big(\frac{\delta _{\mathit{l}}}{\delta \phi
^{C}}\frac{\delta \Gamma _{M}}{\delta \phi _{A}^{\ast }}\Big)\frac{\delta _{%
\mathit{l}}}{\delta \phi ^{B}}\texttt{ and }\varepsilon _{A}\equiv
\varepsilon (\phi ^{A}).  \label{FAdef}
\end{eqnarray}%
On the extremals, $\big(J_{A},\delta \Gamma _{M}/\delta \phi ^{A}\big)=0$,
respectively, for $Z_{M},\Gamma _{M}$, the corresponding variations are
vanishing, $\big(\delta Z_{M}\Big|{}_{J=0},\delta \Gamma _{M}\Big|{}_{\Gamma
_{M{}A}=0}\big)=0$, provided that the gauge variation $\psi \rightarrow \psi
+\delta \psi $ affects not only the BRST exact part of the action, changed
by the term $s\delta \psi $, but also the functional, $M$, $\delta M=-(sM)%
\frac{\imath }{\hbar }\delta \psi $, which was shown in \cite{Reshetnyak}.
Now, we are in a position to apply these results to a special choice of the
BRST symmetry breaking term, when $M=H(A)$.

\section{Gauge-independence in the Gribov--Zwanziger Theory}

\label{GZtheory}

In the case of the GZ theory, the Gribov horizon functional $H(A)$ in the
Landau gauge $\psi _{0}(\phi )$, determined by Eq. (\ref{rxig}) for $\xi =0$
\cite{Zwanziger2} and its Slavnov variation in a non-local formulation read
\begin{eqnarray}
&&H(A)=\gamma ^{2}\,\big(f^{abc}A_{\mu }^{b}(K^{-1})^{ad}f^{dec}A^{e\mu }\
+\ D(N^{2}{-}1)\big)\ ,\mathtt{for}(K^{-1})^{ad}(K)^{db}=\delta ^{ab}
\label{sM} \\
&&\hspace{-0.5em}sH=\gamma ^{2}f^{abc}f^{cde}\bigl[2D_{\mu
}^{bq}C^{q}(K^{-1})^{ad}-f^{mpn}A_{\mu
}^{b}(K^{-1})^{am}K^{pq}C^{q}(K^{-1})^{nd}\bigr]A^{e\mu },
\end{eqnarray}%
with a thermodynamic Gribov parameter, $\gamma $, determined in a
self-consistent way, by using the gap equation $\frac{\partial }{\partial
\gamma }\left( \frac{\hbar }{\text{i}}\,\mbox{ln}\,Z_{H}(0,0)\right) =0$.
The Ward identities for $Z_{H}(J,0)$ and EA, $\Gamma _{H}(\phi ,0)$, are
readily obtained from Eqs. (\ref{wardZM}) and (\ref{WIGammaBV}) as follows:
\begin{eqnarray}
&&  J_{A}\langle s\phi^A\rangle  +\langle H_{\mu a} sA^{\mu a}\rangle  =0,\texttt{ where } \langle \mathcal{O}\rangle = Z^{-1}_{H}(J,0) \mathcal{O}\big({\textstyle\frac{\hbar }{\imath}}{%
\textstyle\frac{\delta }{\delta J}}\big)Z_{H}(J,0), \label{WIZGZ}
  \\
   && \frac{\delta \Gamma_H(\phi,0)%
}{\delta \phi ^{A}} s \phi^A(\widehat{\phi}) \ ={\widehat{H}}_{\mu a}\cdot sA^{\mu a}(\widehat{\phi})\texttt{ where } H_{\mu a}(A) = \frac{\delta H}{\delta A^{\mu a}},  \label{WIGGZ}
\end{eqnarray}%
with the use of the notation of Section~\ref{YMtheory}.

Eqs. (\ref{WIZGZ}), (\ref{WIGGZ}), together with the representations (\ref%
{varZMlin}), (\ref{varGMlin}), (\ref{FAdef}), permit one to obtain the
variations for $Z_{H}(J,0)$ and $\Gamma _{H}(\phi ,0)$ as
\begin{eqnarray}
&&\delta Z_{H}=\left( \frac{\imath }{\hbar }\right) ^{2}\Big(J_{A}\langle
(s\phi ^{A})\delta \psi \rangle +\langle H_{\mu a}(sA^{\mu a})\delta \psi
\rangle +\frac{\hbar }{\imath }\langle \delta H\rangle \Big)Z_{H},
\label{varZHlin} \\
&&\delta \Gamma _{H}\ =\frac{\delta \Gamma _{H}}{\delta \phi ^{A}}{\widehat{F%
}}_{H}^{A}\,\langle \delta \Psi \rangle \big|{}_{\phi ^{\ast }=0}\ -\ {%
\widehat{H}}_{\mu {}a}{\widehat{F}}_{H}^{\mu {}a}\widehat{\delta \psi }\big|%
{}_{\phi ^{\ast }=0}\ +\ \widehat{\delta H}\ .  \label{varGHlin}
\end{eqnarray}%
The gauge variation $\delta G(A)$ induced by the variation $\delta \psi $ of
a functional $G(A)$ defined in the configuration space of the YM fields can
be presented by the gauge transformation with parameters $\xi ^{b}$
constructing from $\delta \psi $, as follows:%
\begin{equation}
\label{gaugevar}
  G(A) \to G(A) + \delta G(A) = G(A) + G_{\mu{}a} D^{\mu{}{ab}} \xi^b,\texttt{ for }\xi^b =  - \frac{\imath}{\hbar} C^b \delta \psi,
\end{equation}%
which was, in fact, shown for the first time with the help of
field-dependent BRST transformations in \cite{ll1}. It is obvious from Eqs. (%
\ref{varZHlin}) and (\ref{varGHlin}) that on the mass shell we have, $\delta
Z_{H}\big|{}_{J=0}=0$, $\delta \Gamma _{H}\big|{}_{\Gamma _{H{}A}=0}=0 $. As
a by-product, the representation for $H(A)$ in a new gauge reference frame, $%
\psi_0 +\delta \psi $, not necessarily related by infinitesimal gauge
variation, follows from Eq. (\ref{gaugevar}) for $G=H$.

\section{A Composite Field Representation for the Gribov Horizon Functional}

\label{CFmethod}

Since unitary cannot be verified explicitly, due to the BRST non-invariance
of the action with the $M(\phi ,\phi ^{\ast })$ [$H(A)$] term for the model
with $Z_{M}$ (\ref{ZM}), in particular, for the GZ model with a non-local $%
H(A)$ (\ref{sM})\footnote{%
For a local representation of the Gribov horizon functional in different
gauges and the study of gauge dependence in Section~\ref{GZtheory}, see \cite%
{ll1}, \cite{Reshetnyak}} we will treat these term as a composite field, $%
\sigma (A)=H(A)$. In doing so, we define a generating functional of Green's
functions with composite fields, by the relation $Z_{L\sigma }=Z_{M}$,%
\begin{eqnarray}\label{Zcf}
Z(J,\phi ^{\ast },L)=Z_{L\sigma }(J,\phi ^{\ast })=\int d\phi \exp \left\{
\frac{\imath }{\hbar }\big(S_{0}(A)+s\psi +\phi _{A}^{\ast }s\phi
^{A}+L_{m}\sigma ^{m}(\phi )+J_{A}\phi ^{A}\big)\right\} .
\end{eqnarray}%
with sources $L_{m}$, $\varepsilon (L_{m})=0$ for $\sigma ^{m}$. Making a
restriction to the case $m=1$, we introduce an EA, $\Gamma (\phi ,\phi
^{\ast },\Sigma )\equiv \Gamma _{\Sigma }$, with a composite field via a
Legendre transformation of $\ln Z_{L\sigma }$ w.r.t. $J,L$ by the relation,
as we follow \cite{compf},
\begin{eqnarray}
&& \hspace{-2.0em}\Gamma_\Sigma = \frac{\hbar}{\imath}\ln Z_{L\sigma} -J\phi- L(\Sigma+\sigma(\phi))\texttt{ for }\phi^A = \frac{\hbar}{\imath}\frac{\delta \ln Z_{L\sigma}}{\delta J_A}\texttt{ and }\Sigma = \frac{\hbar}{\imath}\frac{\delta \ln Z_{L\sigma}}{\delta \sigma}-\sigma(\phi),\label{EAcf}\\
&& \hspace{-2.0em} \texttt{ where }(J_A,L) =  \left(- \frac{\delta\Gamma_\Sigma}{\delta \phi^A}+  \frac{\delta\Gamma_\Sigma}{\delta \Sigma} \frac{\delta\sigma(\phi)}{\delta \phi^A}, - \frac{\delta\Gamma_\Sigma}{\delta \Sigma} \right)\equiv \mathcal{N}_\alpha\texttt{ and }\Phi^\alpha \equiv (\phi^A,\Sigma). \label{Nphi}
 \end{eqnarray}%
Note that the tree approximation for $\Gamma _{\Sigma }$ in the loop
expansion $\Gamma _{\Sigma }=\sum_{n}\hbar \Gamma _{\Sigma }^{(n)}$
coincides with the FP action, $\Gamma _{\Sigma }^{(0)}=S_{0}+s\psi +\phi
^{\ast }s\phi $, providing the influence of the composite field beyond the
tree level. Secondly, the boundary conditions for $L=0$, $Z(J,\phi ^{\ast
},0)=Z_{0}(J,\phi ^{\ast })$, and for $L=1$, $Z(J,\phi ^{\ast
},1)=Z_{H}(J,\phi ^{\ast })$, are fulfilled.

The Ward identities for $Z_{L\sigma }(J,0)$ and $\Gamma _{\Sigma }$ assume
the form, inherited from (\ref{WIZGZ}), (\ref{WIGGZ}),
\begin{eqnarray}
&&  J_{A}\langle s\phi^A\rangle_{L}  +L\langle H_{\mu a} sA^{\mu a}\rangle_{L}  =0,\texttt{ for } \langle \mathcal{O}\rangle_L = Z^{-1}_{L\sigma}(J,0) \mathcal{O}\big({\textstyle\frac{\hbar }{\imath}}{%
\textstyle\frac{\delta }{\delta J}}\big)Z_{L\sigma}(J,0), \label{WIZc}
  \\
   && \frac{\delta \Gamma(\phi,0,\Sigma)%
}{\delta \phi ^{A}} s \phi^A(\widehat{\phi}) \ = -  \frac{\delta \Gamma(\phi,0,\Sigma)%
}{\delta \Sigma}\Big( {\widehat{H}}_{\mu a} - {H}_{\mu a}(\phi) \Big)\cdot sA^{\mu a}(\widehat{\phi}).  \label{WIGc}
\end{eqnarray}%
Again, with the use of these relations the study of gauge dependence for $%
Z_{L\sigma }(J,0)$ and $\Gamma _{\Sigma }$ looks as follows:
\begin{eqnarray}
&& \delta Z_{L\sigma} =  \left(\frac{\imath}{\hbar}\right)^2\Big(J_{A}\langle (s\phi^A) \delta\psi\rangle_L  + L\Big\{\langle H_{\mu a} (sA^{\mu a}) \delta\psi \rangle_L  +\frac{\hbar}{\imath} \langle\delta H\rangle_L\Big\} \Big)Z_{L\sigma} ,\label{varZclin} \\
&&
\delta\Gamma_\Sigma \ = \frac{\delta\Gamma_\Sigma}{\delta\phi^A}
{\widehat F}_{L\sigma}^A\,\langle\delta\Psi\rangle\big|{}_{\phi^*=0}\ -\ \frac{\delta\Gamma_\Sigma}{\delta\Sigma}\Big\{\widehat{\delta
H}\ -\ \big({\widehat
H}_{\mu{}a}-
H_{\mu{}a}(A)\big){\widehat F}_{L\sigma}^{\mu{}a} \widehat{\delta\psi}\big|{}_{\phi^*=0} \Big\}\   \label{varGclin}\\
&&\texttt{ for } {\widehat F}_{L\sigma}^{\mu{}a} \equiv  -\frac{\delta}{\delta A^*_{\mu{}a}}\ +\
(-1)^{\varepsilon_\beta} (K^{''-1})^{\beta \alpha}\Big(\frac{\delta_{\it
l}}{\delta\Phi^\alpha}\frac {\delta
\Gamma_\Sigma}{\delta A^{*}_{\mu{}a}}\Big)\frac{\delta_{\it l}
}{\delta\Phi^\beta}\texttt{ and }(K^{''})_{\beta \alpha} = \frac{\delta\mathcal{N}_\alpha}{\delta \Phi^\beta},
\end{eqnarray}%
where we have used Eqs. (\ref{Zcf})--(\ref{Nphi}) and Ward identities (\ref%
{WIZc}), (\ref{WIGc}). Note that in the above representations for $\delta
Z_{L\sigma }$, $\delta \Gamma _{\Sigma }$ we have taken into account the
variation of the composite field itself (cf. \cite{LO}, \cite{LOR}).

We finally state that on the mass shell determined by $J_{A}=0$ and $(\delta
\Gamma _{\Sigma })/(\delta \phi ^{A}=0$, in view of the transformation of $%
H(A)$ under a gauge variation $\delta \psi $, the generating functional $%
Z_{L\sigma }(J,0)$ and the EA, $\Gamma _{\Sigma }(\phi ,0\Sigma )$,
respectively, do not depend on a choice of the gauge for any values of the
source $L$ and the additional extremal $(\delta \Gamma _{\Sigma })/(\delta
\Sigma $, thus providing the gauge independence of the $S$-matrix.

Concluding, notice that one may use the latter concept for the Gribov
horizon functional as a composite field in the study of renormalizability and unitarity of the GZ
theory starting from a specific gauge, e.g. Landau gauge,  in order to investigate the consistency of the
composite field approach with soft BRST symmetry breaking in YM theories.

\vspace{-1ex}

\paragraph{Acknowledgements}

The author is thankful
to the organizers of the International Workshop SQS'13 for the
hospitality,  to V.P. Gusynin, R. Metsaev, P. Moshin  for interest  and discussions, to I.V. Tyutin for discussion of the unitarity problem, to Sh. Gongyo for useful correspondence. The study was supported by the RFBR grant   Nr. 12-02-00121 and by LRSS grant Nr. 88.2014.2.

\end{document}